An Algebraic Derivation of the Standing Wave Problem

Thomas A. Dooling
Department of Chemistry and Physics
University of North Carolina Pembroke
Pembroke, NC 28372-1510
tom.dooling@uncp.edu

William D. Brandon
Department of Chemistry and Physics
University of North Carolina Pembroke
Pembroke, NC 28372-1510
william.brandon@uncp.edu

Abstract:
The standing wave solution on an idealized mass spring system can be found using straight forward algebra. The solution is found when this system makes "jump-rope" like rotations around an axis. The standing wave forms a constant shape in a radial direction using the centripetal force condition. The wave is projected back onto the x-y plane to get the planar time dependent solutions. The allowed frequencies are found for a discrete system as well as a continuous system.

One may consider a system of springs and masses as forming an idealized stretched string. Solutions of this problem are often used as a starting point for understanding standing waves on a string[1]. When transverse modes of oscillation are combined, we find that the motion of string can stabilize into circular movements, like a jump-rope[2]. In fact, it is commonly observed that a harmonically driven string prefers circular motion to simple transverse oscillation when tuned to "resonance", unless care is taken to drive the string with low amplitude. This whirling motion, often witnessed in the introductory physics laboratory[3], has since been described[4] and measured[5] as a coupling of the directly driven transverse mode and an additional parametrically excited orthogonal mode. However, due to the complexity of the problem, any discussion of circular motion is overlooked in all the standard general physics textbooks we have come across. In contrast to the usual development offered in these texts, one can start from the jump-rope condition, as justified by observation, and then the standing wave solution for the transverse modes can be found in a straightforward algebraic manner using vectorial projection. Therefore, this treatment also lends itself quite naturally as starting point to mathematically similar systems such as circularly polarized light.

Assume a system of springs and masses is fixed on both ends. The masses lay in a plane that rotates about the z axis with an angular frequency $w$ (Fig. 1). Each mass maintains a fixed radial distance (r) and fixed position on the z axis (z). In other words, the string is in "equilibrium" in a coordinate system rotating with angular frequency ($w$).

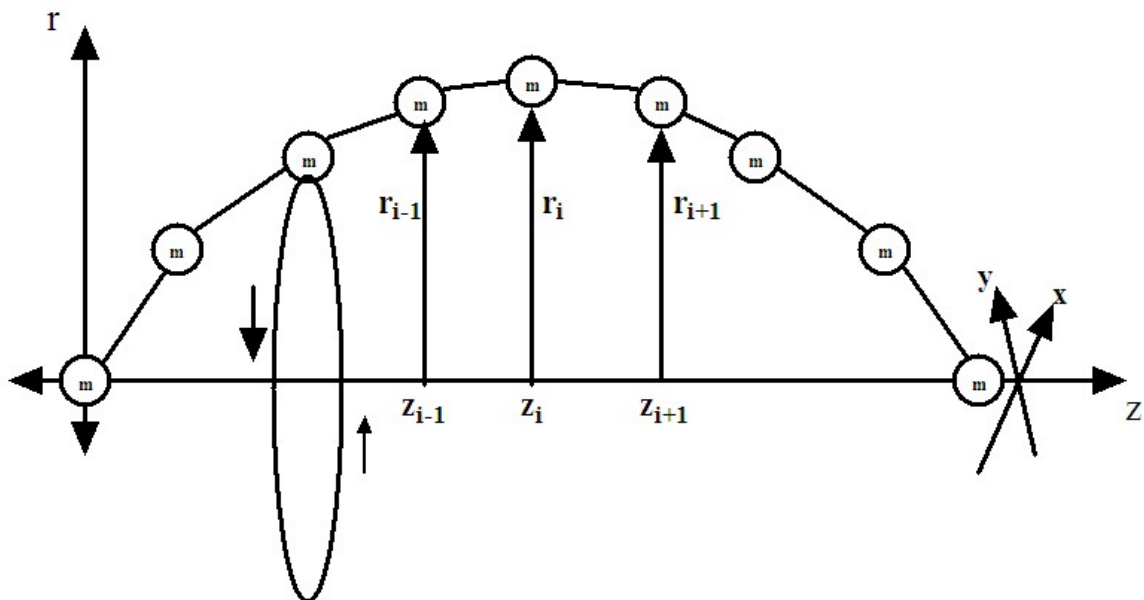

**Fig. 1. N particles and (N-1) springs. The first particle is labeled as (0) and the last particle as (N-1).**

Each mass experiences a force projected onto the z direction and the radial direction (Fig. 2.).

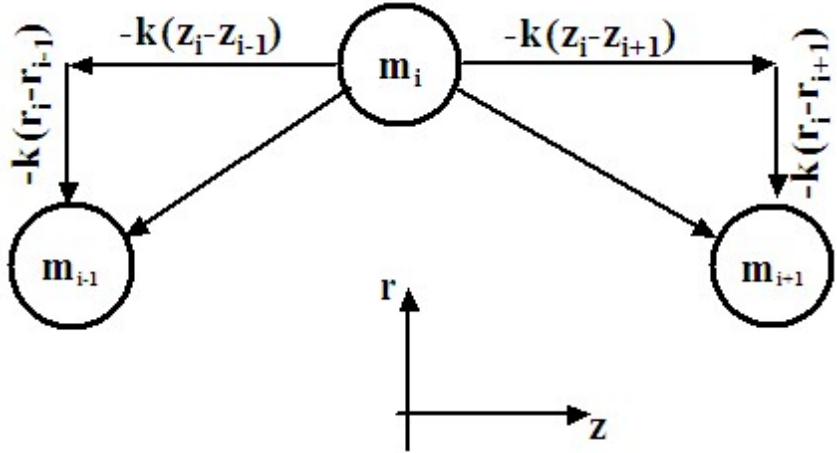

**Fig. 2. The spring force components in the linear and radial directions.**

In the radial direction, each particle is held in place by a centripetal force[6], expressed by the equations,

$$-k(r_i - r_{i-1}) - k(r_i - r_{i+1}) = -m r_i w^2 \qquad (1)$$

Rearranging we have,
$$w_0^2 \left[ (r_i - r_{i-1}) + (r_i - r_{i+1}) \right] = w^2 r_i \quad . \qquad (2)$$
where the natural frequency is given by $w_0 = (k/m)^{1/2}$. This is similar to the linear difference equation developed in Marion[7]. We have arrived at this equation using the centripetal force condition, bypassing a more involved calculus based derivation.

In order to develop a solution to Eq. (2), we note that the forces in the (z) direction must sum to zero as shown below,

$$\begin{aligned}
-k(z_i - z_{i-1}) - k(z_i - z_{i+1}) &= 0 \quad &(a) \\
-(z_i - z_{i-1}) &= (z_i - z_{i+1}) \quad &(b) \\
(z_i - z_{i-1}) &= (z_{i+1} - z_i) \quad &(c)
\end{aligned} \qquad (3)$$

Eq. (3.c) shows that for stable, circular "jump rope" like motion, $\Delta z$ = constant. When each particle position on the z axis satisfies this condition we are assuming an ideal string with no longitudinal compression. Eq. (3.c) implies

$$z_i = i \Delta z \, , \, i = 0, 1, 2 ... \qquad (4)$$

Assuming a solution to the difference equation, Eq. (2), given by

$$r_i = \sin(\gamma z_i) = \sin(\gamma i \Delta z), \tag{5}$$

where $\gamma$ is a constant to be determined later and substituting Eq. (5) into Eq. (2) yields

$$\omega_0^2 \left\{ [\sin(\gamma i \Delta z) - \sin(\gamma(i+1)\Delta z)] + [\sin(\gamma i \Delta z) - \sin(\gamma(i-1)\Delta z)] \right\} = \omega^2 \sin(\gamma i \Delta z) \tag{6}$$

Since $\sin(A \pm B) = \sin(A)\cos(B) \pm \cos(A)\sin(B)$ the arguments of the $\sin(\gamma(i \pm 1)\Delta z)$ terms can be expanded and divided out so that Eq. (6) reduces to

$$\omega_0^2 \{2 - 2\cos(\gamma \Delta z)\} = \omega^2. \tag{7}$$

Finally, the angular frequency can be expressed as,

$$w = 2w_o \sqrt{\frac{(1-\cos(\gamma \Delta z))}{2}} = 2w_o \sin\left(\frac{\gamma \Delta z}{2}\right). \tag{8}$$

Now, since the string is constrained on both ends the following condition holds true

$$\sin(\gamma L) = \sin(\gamma (N-1)\Delta z) = 0. \tag{9}$$

As a result,

$$\gamma(N-1)\Delta z = j\pi$$
$$\gamma \Delta z = \frac{j\pi}{(N-1)} \quad , \text{ where j=1,2,3...} \tag{10}$$

Then,

$$w = 2w_o \sin\left(\frac{\gamma \Delta z}{2}\right) = 2w_o \sin\left(\frac{j\pi}{2(N-1)}\right) \tag{11}$$

Since, ($\Delta z(N-1) = L$), a solution for the shape of the string follows,

$$r_i = \sin(\gamma z_i) = \sin\left(z_i \frac{j\pi}{\Delta z(N-1)}\right) = \sin\left(z_i \frac{j\pi}{L}\right) \tag{12}$$

where, i equals the particle number.

If we started with the general problem of an idealized mass spring system, the equations of motion for the x and y directions would be, by Newton's 2nd Law;

$$-k(x_i - x_{i-1}) - k(x_i - x_{i+1}) = ma_{xi} \quad (a)$$
$$-k(y_i - y_{i-1}) - k(y_i - y_{i+1}) = ma_{yi} \quad (b) \tag{13}$$

Both the x and y directions are perpendicular to the z axis along which the string is stretched. Unlike the equations developed from a more general treatment displaying coupled modes, this idealized system has fully decoupled equations. So, if we project the motion of the system back into the x and y coordinates we have automatically found solutions for the x and y motion which are independent from each other.

The x and y solutions follow from the vectorial projection of $r_i$ back onto these axes is,

$$x_i = A_x r_i \cos(wt) \quad (a)$$
$$y_i = A_y r_i \sin(wt) \quad (b) \tag{14}$$

Since the x and y motions are decoupled, they can each be given their own arbitrary amplitude.

There are different methods for extrapolating the above solutions to the continuous case. One involves conversion of the difference equation into a second order differential equation[8]. We will proceed with another method that first solves for the angular frequency of a continuous string[9] in the following manner. Allow the number of particles to approach infinity, ($N \to \infty$). Then the Sine term will be equal to its own argument as the argument approaches zero. That is,

$$w = 2w_o \sin\left(\frac{j\pi}{2(N-1)}\right) \xrightarrow{N \to \infty} 2w_o \left(\frac{j\pi}{2(N-1)}\right) \tag{15}$$

and

$$w = 2w_o \left(\frac{j\pi}{2(N-1)}\right) = \frac{w_o j\pi}{N-1} = \sqrt{\frac{k}{m}} \frac{j\pi\Delta z}{L}. \tag{16}$$

Since ($\Delta z$ = constant), the mass of the system is distributed evenly along the z axis. The linear mass density of the string is equal to is mass divided by its length. ($\rho = L/M$) Then, for a large number of masses which add to M, the individual masses are,

$$m = \rho \Delta z. \tag{17}$$

The spring constant ($k$) is determined by the rest tension in the string and the separation between the masses;

$$T_r = k\Delta z \tag{18}$$

where $T_r$ is the rest tension before transverse displacement and $\rho$ is the linear mass density.

Substituting for the mass and spring constant, the standard solution for the angular frequencies of a vibrating string are obtained.

$$w = \sqrt{\frac{k}{m}} \frac{j\pi\Delta z}{L} = \sqrt{\frac{T_r}{\rho\Delta z\Delta z}} \frac{j\pi\Delta z}{L} = j\sqrt{\frac{T_r}{\rho}} \frac{\pi}{L} , j = 1,2,3... \quad (19)$$

The standing wave frequencies become,

$$f = \frac{\omega}{2\pi} = \frac{j}{2L}\sqrt{\frac{T_r}{\rho}}, \quad (20)$$

Finally, dropping the (i) subscript in the continuous limit, we obtain the solution for standing waves on an idealized string is,

$$r = \sin\left(z\frac{j\pi}{L}\right).$$

(21)

and the standard solution follows[10],

$$x = A_x \sin\left(z\frac{j\pi}{L}\right)\cos(wt) \quad (a)$$

$$y = A_y \sin\left(z\frac{j\pi}{L}\right)\sin(wt) \quad (b)$$

(22)

The goal of this paper was to show that the standard solution to a vibrating string could be derived using simple algebra. This non-calculus solution makes the problem more accessible to algebra based physics courses and provides a little more insight, into the nature of string vibrations. This naturally points towards other phenomenon like polarization, while addressing the more realistic "whirling" behavior likely to be witnessed, even in the introductory laboratory course.